# Isotropic superconductivity in pressurized trilayer nickelate La$_4$Ni$_3$O$_{10-\delta}$

Di Peng[1,#], Yaolong Bian[2,#], Zhenfang Xing[3,#], Lixing Chen[4#], Jiaqiang Cai[2], Tao Luo[3], Fujun Lan[3], Yuxin Liu[3], Yinghao Zhu[4], Enkang Zhang[4], Zhaosheng Wang[2], Yuping Sun[2], Yuzhu Wang[5], Xingya Wang[5], Chenyue Wang[5], Yuqi Yang[6], Yanping Yang[3], Hongliang Dong[3], Hongbo Lou[3], Zhidan Zeng[3], Zhi Zeng[7], Mingliang Tian[2], Jun Zhao[4,8\*], Qiaoshi Zeng[1,3,\*], Jinglei Zhang[2,\*], Ho-kwang Mao[1,3]

[1] Shanghai Key Laboratory of Material Frontiers Research in Extreme Environments (MFree), Institute for Shanghai Advanced Research in Physical Sciences (SHARPS), Shanghai, China
[2] Anhui Province Key Laboratory of Condensed Matter Physics at Extreme Conditions, High Magnetic Field Laboratory, HFIPS, Chinese Academy of Sciences, Hefei, China
[3] Center for High Pressure Science and Technology Advanced Research, Shanghai, China
[4] State Key Laboratory of Surface Physics and Department of Physics, Fudan University, Shanghai, China
[5] Shanghai Synchrotron Radiation Facility, Shanghai Advanced Research Institute, Chinese Academy of Sciences, Shanghai, China
[6] Human Institute, ShanghaiTech University, Shanghai, China
[7] Key Laboratory of Materials Physics, Anhui Key Laboratory of Nanomaterials and Nanotechnology, Institute of Solid State Physics, HFIPS, Chinese Academy of Sciences, Hefei, China
[8] Institute of Nanoelectronics and Quantum Computing, Fudan University, Shanghai, China
[#] These authors contribute equally to this work.
\* To whom correspondence should be addressed: zengqs@hpstar.ac.cn, zhaoj@fudan.edu.cn, zhangjinglei@hmfl.ac.cn.

**Evidence of superconductivity (SC) has recently been reported in pressurized La$_3$Ni$_2$O$_{7-\delta}$ and La$_4$Ni$_3$O$_{10-\delta}$, providing a new platform to explore high-temperature superconductivity. However, while zero resistance state has been observed, experimental characterization of the superconducting properties of pressurized nickelates is still limited and experimentally challenging. Here, we present the first full temperature dependence of the upper critical field ($\mu_0 H_{c2}$) measurement in La$_4$Ni$_3$O$_{10-\delta}$ single crystal, achieved by combining high magnetic field and high-pressure techniques. Remarkably, the $\mu_0 H_{c2}$ of La$_4$Ni$_3$O$_{10-\delta}$ is nearly isotropic, with the anisotropic parameter γ monotonically increasing from 1.4 near $T_c$ to 1 at lower temperatures. By analyzing the $\mu_0 H_{c2}^\perp(T)$ and $\mu_0 H_{c2}^\parallel(T)$ using the two-band model, we uncover that the anisotropic diffusivity of the bands, primarily originating from $d_{z^2}$ and $d_{x^2-y^2}$ orbitals, is well compensated, resulting in an unusually isotropic superconducting state. These findings provide critical experimental evidence that underscores the significant role of the $d_{z^2}$ orbital in enabling superconductivity in pressurized Ruddlesden-Popper nickelates.**

**Main**

The recent discovery of superconductivity (SC) in Ruddlesden-Popper (RP) nickelates has attracted worldwide interest in searching for new high-temperature superconductors. The first breakthrough along this line is the observation of superconductivity in the bilayer RP nickelate $La_3Ni_2O_{7-\delta}$ with the critical temperature ($T_c$) as high as 80 K at pressures above 14 GPa[1]. Subsequent studies revealed zero-resistance states under hydrostatic pressure[2-4]. However, the superconducting volume fraction for the single-crystalline $La_3Ni_2O_{7-\delta}$ remained low[3,5]. The unambiguous bulk superconductivity with the $T_c$ up to 30 K was reported in the single crystal of trilayer RP nickelate $La_4Ni_3O_{10-\delta}$ under pressure[6], although the $T_c$ and superconducting volume fraction are dependent on sample quality and pressure conditions[7,8]. Very recently, superconductivity was reported in bilayer nickelate ultrathin films with substrate confinement[9-11]. Both $La_3Ni_2O_{7-\delta}$ and $La_4Ni_3O_{10-\delta}$ host a quasi-2D crystal structure, with an approximate unit cell comprising $NiO_2$ layers that is isostructural with the $CuO_2$ plane in the superconducting cuprates[1,6]. However, the presence of apical oxygens in RP nickelates leads to unique electronic configurations ($d^{7.5}$ for $La_3Ni_2O_7$ and $d^{7.33}$ for $La_4Ni_3O_{10}$), significantly different from the $3d^9$ configuration in cuprates[1,6,12]. Numerous theoretical studies suggest that interlayer coupling between Ni $3d_{z^2}$ orbitals and apical oxygen $p$ orbitals at high pressures leads to partially occupied $3d_{x^2-y^2}$ orbitals, which has sparked debate regarding the precise roles of the $d_{z^2}$ and $d_{x^2-y^2}$ orbitals in the electron pairing mechanisms[12-36]. Experimental studies, especially those that can directly probe the superconducting characteristics of RP nickelates under high pressure, are critical to address this issue.

The upper critical field ($\mu_0 H_{c2}$) is a fundamental characteristic of a superconductor, which is crucial for understanding the superconducting pairing mechanism as well as for potential applications. Especially the anisotropy of $\mu_0 H_{c2}$ is an effective tool to reveal the topology of the underlying electronic structure for layer superconductors. For instance, highly anisotropic $\mu_0 H_{c2}$ in cuprates reflects the two-dimensional (2D) nature of their crystal and electronic structures[37,38]. In contrast, the weak anisotropy of $\mu_0 H_{c2}$ in most iron-based SCs is conjectured to originate from the combination of a more three-dimensional (3D) Fermi surface and the multiband effect[39,40]. Thus, directly determining $\mu_0 H_{c2}$ and its anisotropy for RP nickelate-based SC are timely important. However, due to the necessity of high pressure to induce superconductivity and a high magnet field to access $\mu_0 H_{c2}(0)$, experimental progress in this direction is markedly hampered. The technical challenge lies in simultaneously achieving hydrostatic pressures exceeding 50 GPa and steady magnetic fields exceeding 30 T while maintaining cryogenic conditions. The bulky dimensions of conventional pressure cells restrict modifications to crystal axes orientations under high pressures, which are essential for conducting anisotropy measurements.

In this study, we present direct measurements of electrical resistivity in single-crystal $La_4Ni_3O_{10-\delta}$ under extreme conditions, with a pressure up to 50.2 GPa and a magnetic field up to 34 T for both out-of-plane ($H \perp ab$) and in-plane ($H // ab$) directions. The temperature–magnetic-field phase diagram of $La_4Ni_3O_{10-\delta}$ is well established. It was found that the upper critical field of La$_4$Ni$_3$O$_{10-\delta}$ nearly exhibits isotropy over the entire temperature range, with the anisotropic ratio $\gamma = \mu_0 H_{c2}^{\parallel}/\mu_0 H_{c2}^{\perp}$ monotonically decreases from 1.4 near $T_c$ to 1 at the lowest measured temperature. The temperature dependence of $\mu_0 H_{c2}(T)$, in both field directions, can be described by the two-band model, allowing us to quantitatively evaluate the diffusivity of each band. Our analysis further demonstrates that the $d_{z^2}$ and $d_{x^2-y^2}$ bands have opposing anisotropies, leading to such isotropic superconductivity in La$_4$Ni$_3$O$_{10-\delta}$.

We first focus on the superconductivity of single-crystal La$_4$Ni$_3$O$_{10-\delta}$ at zero-field. The electrical resistance measurements under pressure were performed using a homemade diamond anvil cell (DAC). Helium was employed as the pressure-transmitting medium to guarantee the best hydrostaticity in the DAC. As shown in the inset of Fig.1, the sample is loaded with the *ab*-plane, which is the cleavage plane of a La$_4$Ni$_3$O$_{10-\delta}$ single crystal, parallel to the diamond anvil culet of the DAC. Synchrotron X-ray single-crystal diffraction measurements demonstrate that the sample we studied maintains its single-crystal structure feature up to 50.2 GPa (Extended Data Fig. 1). Figure 1a displays the resistance as a function of temperature at 50.2 GPa. A clear superconducting transition is observed in resistance, with an onset at 21.8 K and a midpoint at 18.5 K, respectively. To verify bulk superconductivity, direct current (DC) magnetic susceptibility was measured with sample (S#2) from the same batch. The bulk $T_c$ ($T_c^{onest}$ =18.6 K and $T_c^{mid}$=14.9 K) determined from the magnetic susceptibility is slightly lower than that determined from the resistance. The estimated superconducting volume fraction of our sample is approximately 86%[6]. The normal state resistance follows a linear temperature dependence up to ~125 K, indicative of strange-metal behavior. Such characteristics had been widely observed in optimally doped cuprates[41,42], certain iron-based superconductors[43,44], and nickelate-based superconductors[2,6,45]. Most importantly, when superconductivity is completely suppressed by a magnetic field of 34 T, the *T*-linear resistance persists to the lowest-measured temperature, implying the presence of a pressure-induced quantum critical point in La$_4$Ni$_3$O$_{10-\delta}$.

Figures 2a and b show the magnetic-field dependence of the electrical resistance for La$_4$Ni$_3$O$_{10-\delta}$ single crystals. In order to estimate the anisotropy of the superconductivity, the measurements were conducted with the magnetic field applied along the out-of-plane (a) and in-plane (b) directions. The superconducting transition gradually shifted to a lower magnetic field with increasing temperature. In the presence of an in-plane magnetic field, the SC transition is broadened significantly at higher temperatures. We noticed that such a broadening behavior is not observed in the *R-T* curve under a magnetic field (Extended Data Fig. 2). Thus, it is likely attributed to a magnetoresistance effect, as we discussed below, instead of thermally activated flux flow in the vortex state as reported in La$_3$Ni$_2$O$_{7-\delta}$[2,3]. Above $T_c$, the magnetoresistance (MR) of La$_4$Ni$_3$O$_{10-\delta}$ displays distinct field-orientation dependence. Specifically, the out-of-plane MR, which exhibits a weak "hump" feature, is quite small, whereas more pronounced positive MR is observed for *H//ab*. Further investigation is needed in the further to clarify such anisotropic MR in its normal state.

Figure 3 illustrates the upper critical fields as a function of temperature for magnetic fields applied perpendicular ($\mu_0 H_{c2}^{\perp}(T)$) and parallel ($\mu_0 H_{c2}^{\parallel}(T)$) to the *ab*-plane. To minimize the effects of magnetoresistance and superconducting fluctuations, the critical fields and critical temperatures are determined from the midpoint of the superconducting transitions. The $\mu_0 H_{c2}^{\perp}(T)$ increases linearly with decreasing temperature. On the contrary, the $\mu_0 H_{c2}^{\parallel}(T)$ shows a convex curvature. The upper critical field reaches approximately 24.4 T at the lowest temperature for both field orientations. The estimated coherence length ($\xi_{\parallel}(1.8\ \text{K}) \approx \xi_{\perp}(1.8\ \text{K}) = 33\ \text{Å}$) is an order of magnitude larger than the thickness of the superconducting NiO$_2$ layers ($\sim 3.7\ \text{Å}$)[6], demonstrating the 3D nature of the superconductivity. It is interesting to note that the nearly isotropic upper critical field can extend across the entire temperature range, as evidenced by the fact that the anisotropic ratio $\gamma = \mu_0 H_{c2}^{\parallel} / \mu_0 H_{c2}^{\perp}$ only varies from 1 (at base temperature) to 1.4 (at near $T_c$). The results are reproducibly observed on another sample (S#4 in Extended Data Figs. 3-5).

In a superconductor, the Cooper pairs can be destroyed by a magnetic field through two primary mechanisms. (i) The orbital pair breaking due to the Lorentz force acting via the charge on the paired electrons, known as the orbital limit[46]. The orbital-limited upper critical field $\mu_0 H_{c2}^{orb}(0)$ for a single-band BCS superconductor is determined by the initial slope of $\mu_0 H_{c2}$ at $T_c$, i.e. $\mu_0 H_{c2}^{orb}(0) = -0.69 T_c \left(\frac{dH_{c2}}{dT}\right)_{|T=T_c}$. (ii) The Pauli paramagnetic pair breaks as a result of the Zeeman effect, which aligns the spins of two electrons with the applied field[47]. The Pauli paramagnetic limiting field for weakly coupled BCS superconductors is given by $\mu_0 H_{c2}^p(0) = 1.86 T_c$, which is usually field-orientation independent. In our case, the initial slope of $\mu_0 H_{c2}$ at $T_c$ is determined as 1.5 T/K and 2.5 T/K for *H*⊥*ab* and *H*//*ab*, respectively. The derived out-of-plane and in-plane $\mu_0 H_{c2}^{orb}(0)$ are 19.2 T and 30.9 T, respectively, both of which are below the Pauli limit of 34.4 T. To comprehensively evaluate the spin-paramagnetic and orbital pair-breaking effects, we fitted the experimental data of $\mu_0 H_{c2}(T)$ with the Werthamer-Helfand-Hohenberg (WHH) model[46]. The fitting parameters of $\alpha$ (Maki parameter) and $\lambda_{so}$ are defined by $\alpha = \sqrt{2} \frac{\mu_0 H_{c2}^{orb}(0)}{\mu_0 H^p(0)}$ and $\lambda_{so} = \frac{1}{3\pi T_c \tau_2}$, where $\frac{1}{\tau_2}$ reflects the spin-orbit scattering rate. As shown in Fig. 3, neither $\mu_0 H_{c2}^{\perp}(T)$ and $\mu_0 H_{c2}^{\parallel}(T)$ can be accurately described by the WHH model when the spin effects are neglected ($\alpha = 0$ and $\lambda_{so} = 0$). The experimental data for $\mu_0 H_{c2}^{\perp}(T)$ lies above the theoretical curve. By adjusting the values of $\alpha$ and $\lambda_{so}$, $\mu_0 H_{c2}^{\parallel}(T)$ can be approximately represented by the theoretical model with the parameters $\alpha = 1.45$ and $\lambda_{so} = 1.08$. However, the small value of Maki parameter indicates that the orbital effect is the dominant pair-breaking mechanism in both field directions. Thus, the scenario of isotropic Pauli-limited superconductivity, as proposed for infinite-layer nickel-based SCs[48], is unlikely to apply to the isotropic upper critical field of La$_4$Ni$_3$O$_{10-\delta}$.

Previous theoretical studies have predicted that superconductivity in pressurized RP nicklates may emerge in several disconnected regions of the Fermi surface[12,33]. Here, we adopt a two-band model to fit the upper critical field of La$_4$Ni$_3$O$_{10-\delta}$[49]. The details of the parameters in the two-band theory are outlined in the extended analysis and discussion. We first focus on the coupling constant $\lambda_{nm}$, emphasizing two scenarios: strong intraband coupling ($\lambda_{12}\lambda_{21} \ll \lambda_{11}\lambda_{22}$) and strong interband coupling ($\lambda_{12}\lambda_{21} \gg \lambda_{11}\lambda_{22}$). The former was discussed in dirty $MgB_2$[49], while the latter is relevant to some iron-based SCs[50]. When the magnetic field is applied in the out-of-plane direction, the fittings in these two cases exhibit a significant difference. The possibility of strong intraband coupling can be ruled out in $La_4Ni_3O_{10-\delta}$. As the calculated $\mu_0H_{c2}^\perp(T)$ with $\lambda_{12}\lambda_{21} \ll \lambda_{11}\lambda_{22}$ exhibits a strong upturn feature at low temperatures, deviating from the measured data (see Extended Data Fig. 6). The discussions in the main text (represented by the solid line in Fig. 3) are based on strong interband coupling, ($\lambda_{11}=\lambda_{22}=0.2$ and $\lambda_{12}=\lambda_{21}=0.7$), which emphasizes the importance of the strong interband coupling for superconductivity in RP nickelates.

Now, we turn to $D_1$ and $D_2$, the diffusivity of each band. The magnitude of $\mu_0H_{c2}(T)$ is highly sensitive to the diffusivity of each band. Specifically, the upper critical field near $T_c$ is primarily determined by the band with the higher diffusivity $D_1$. The band with lower diffusivity $D_2$ dominates the behavior of $\mu_0H_{c2}(T)$ at low temperatures[49]. Since the two-band model we adopted here ignores the Pauli paramagnetic effect, the fit for $\mu_0H_{c2}^\parallel(T)$ exhibits some deviations. Nevertheless, even taking these fitting uncertainties into account, we can still draw qualitative conclusions that are robust to the fitting details. Two sets of $D_1$ and $D_2$ are obtained by fitting $\mu_0H_{c2}^\parallel(T)$ and $\mu_0H_{c2}^\perp(T)$ (See Extended Data Table 1). DFT calculations reveal that five bands ($\alpha$, $\beta_1$, $\beta_2$, $\gamma$, and $\delta$) cross the Fermi level in pressurized $La_4Ni_3O_{10-\delta}$[33]. The $d_{x^2-y^2}$ orbitals play a dominant role in the formation of the nearly cylindrical two-dimensional pockets ($\alpha$, $\beta_1$, and $\beta_2$), while the other pockets ($\gamma$ and $\delta$) uniquely arise from the $d_{z^2}$ orbital. For simplicity, the diffusivity of the former group is defined as $D_{d_{x^2-y^2}}$, while that of the $d_{z^2}$ orbital-derived pockets is defined as $D_{d_{z^2}}$. Generally, the $d_{x^2-y^2}$ electrons are more itinerant within $ab$-plane, leading to high in-plane diffusivity, $D_{d_{x^2-y^2}}^\parallel > D_{d_{x^2-y^2}}^\perp$. On the other hand, the $d_{z^2}$ electrons have opposing anisotropy of diffusivity where $D_{d_{z^2}}^\parallel < D_{d_{z^2}}^\perp$. Based on these characteristics, we can assign the fitted diffusivity in each field direction to the corresponding bands, which gives $D_{d_{x^2-y^2}}^\parallel = 1.2\ cm^2/sec$, $D_{d_{x^2-y^2}}^\perp = 0.39\ cm^2/sec$, $D_{d_{z^2}}^\parallel = 0.09\ cm^2/sec$, and $D_{d_{z^2}}^\perp = 0.71\ cm^2/sec$, respectively.

At ambient pressure, the Fermi surface of RP nicklates is primarily derived from the $d_{x^2-y^2}$ orbital, which bears notable similarities to cuprates[51]. The emergence of superconductivity under pressure is concomitant with the structural transition, raising a key question of whether the $d_{z^2}$ dominant bands intersect the Fermi level. If the electronic structure is mainly characterized by the $d_{x^2-y^2}$ orbitals, the upper critical field would typically be highly anisotropic. Clearly, this is not the case for $La_4Ni_3O_{10-\delta}$, where both $d_{z^2}$ and $d_{x^2-y^2}$ orbitals must be taken into account to understand the isotropic upper critical field. As schematically illustrated in Fig.3, the behavior of $\mu_0H_{c2}^\perp(T)$ in the high- and low-temperature regimes is primarily determined by $D_{d_{x^2-y^2}}^\parallel$ and $D_{d_{z^2}}^\parallel$, respectively.

In contrast, the behavior of $\mu_0 H_{c2}^{\parallel}(T)$ across these temperature ranges is predominantly controlled by $D_{d_{z^2}}^{\perp}$ and $D_{d_{x^2-y^2}}^{\perp}$. Recently, highly anisotropic superconductivity has been observed in bilayer nickelate ultrathin films[9-11]. It should be noticed that their superconductivity resides in a regime where the coherence length is comparable to the thickness of the thin film. In this context, when H//ab, superconductivity is suppressed at the Clogston-Chandrasekhar limit (CCL), rather than through the orbital mechanism. Thus, it is hard to estimate the intrinsic role of $d_{z^2}$ orbital by analyzing the $\mu_0 H_{c2}^{\parallel}$ of those bilayer nickelate ultrathin films. The nearly isotropic superconductivity of $La_4Ni_3O_{10-\delta}$ is reminiscent of the behavior observed in optimally doped iron-based superconductors, such as $Ba_{0.6}K_{0.4}Fe_2As_2$, where the anisotropic parameter is approximately 2 near $T_c$ and approaches 1 at low temperatures[40]. In the case of iron-based superconductors, interband interactions associated with multiple orbitals also play a significant role.

The $T_c$ of trilayer $La_4Ni_3O_{10-\delta}$ is lower than that of bilayer $La_3Ni_2O_{7-\delta}$. This contrasts with cuprates, for which the highest $T_c$ is achieved in trilayer systems. Such a discrepancy implies the presence of distinct interlayer interaction mechanisms between these two compounds. In the case of $La_4Ni_3O_{10-\delta}$, the diffusivity of the $d_{z^2}$ bands, estimated from $\mu_0 H_{c2}$, can be eight times greater along the out-of-plane direction compared to the *ab* plane ($D_{d_{z^2}}^{\perp}/D_{d_{z^2}}^{\parallel} \sim 8$). It suggests that the bonding state of $d_{z^2}$ orbital always has a Wannier function that is elongated along the *c*-axis all over the layers. It is consistent with the theoretical prediction that the increase in the $NiO_2$ layer would weaken the correlation effect, which leads to the lower $T_c$ in $La_4Ni_3O_{10-\delta}$[35]. Unlike $La_4Ni_3O_{10-\delta}$, the superconducting transition of $La_3Ni_2O_{7-\delta}$ becomes broader with increasing magnetic field[2,3]. Considering the fact that a similar feature, which is the consequence of the formation of a vortex-liquid state, is often observed in the superconductor with large anisotropic $\mu_0 H_{c2}$, including the cuprates and some iron based SCs[52-54]. It implies that the $\mu_0 H_{c2}$ of $La_3Ni_2O_{7-\delta}$ would be more anisotropic compared to $La_4Ni_3O_{10-\delta}$. Nonetheless, the direct investigation of the upper critical fields in pressurized $La_3Ni_2O_{7-\delta}$, particularly systematic studies of the relationship between their anisotropy and critical temperature, is crucial for understanding the superconductive nickel-based materials.

In summary, we have determined the complete temperature-magnetic field phase diagram for the superconductor $La_4Ni_3O_{10-\delta}$. The upper critical field of $La_4Ni_3O_{10-\delta}$ is nearly isotropic in the entire temperature range of superconductivity we studied. The observed shapes of $\mu_0 H_{c2}$ in both field directions can be described by the two-band model. Our quantitative analysis suggests that the isotropic nature of $\mu_0 H_{c2}$ in $La_4Ni_3O_{10-\delta}$ arises from an effective balance between the anisotropic diffusivity of the bands contributed by $d_{z^2}$ and $d_{x^2-y^2}$ orbitals. These upper critical field measurements provide evidence supporting the involvement of the $d_{z^2}$ band in trilayer RP nickelates under pressure and highlight its crucial role in the emergency of superconductivity.

## Methods
### Sample preparations

The precursor powder for the $La_4Ni_3O_{10-\delta}$ compound was prepared using the conventional solid-state reaction method. Single crystals were further grown using a vertical optical-image floating-zone furnace (Model HKZ, SciDre). Details of the crystal growth procedure are given in Ref [6].

### Transport measurements under pressure.

Measurements under pressure were carried out utilizing a diamond anvil cell (DAC) with 240 um culets. The lamella was carved from a single crystal by using focused-ion-beam (FIB) milling and loaded with the *ab*-plane parallel to the diamond anvil culet of the DAC. In all the high-pressure measurements conducted in this work, helium was employed as the pressure-transmitting medium to ensure optimal hydrostatic pressure conditions. Pressure calibration was accomplished by monitoring the shift of the fluorescence peak from ruby balls loaded along with the sample. The temperature dependence of the resistance was measured in a Physical Property Measurement System (PPMS Quantum Design) and a $He^4$ cryostat in the high magnetic field facility. The DC magnetic susceptibility measurement under pressure was conducted using a custom-built beryllium-copper alloy miniature DAC. The measure was performed using a Magnetic Property Measurement system (MPMS3 Quantum design).

### Synchrotron X-ray single crystal diffraction measurements

In situ high-pressure single-crystal synchrotron X-ray diffraction measurements were performed with a wavelength of 0.4834 Å at the beamline 17UM of Shanghai Synchrotron Radiation Facility, China. The measurements were carried out employing a symmetric diamond anvil cell with diamond culet sizes of ~300 μm and a rhenium gasket. The single-crystal sample was prepared by slicing it into a circular disk with a thickness of 6 μm and a diameter of 30 μm using focused ion beam (FIB). The sample was then loaded into a sample chamber with a thickness of approximately 40 μm and a diameter of 180 μm drilled in the pre-indented rhenium gasket.

### Transport measurements under a high magnetic field.

The high-field transport properties measurements up to 35 T were performed at the Chinese High Magnetic Field Laboratory at Hefei using a resistive water-cooled magnet (WM#5). The DAC was mounted on the homemade spring-based rotator probe. The angle is determined by the Hall sensor fixed on the DAC. The resistivity was measured using the AC lock-in technique, where the current was generated by a current source (Keithely 6221), and the voltage was measured with a lock-in amplifier (SR 830).

## Acknowledgements
We acknowledge the helpful discussions with H.Q. Yuan, L. Jiao, and Z.J. Xiang. This work was supported by the National Key R&D Program of China (2021YFA0718900), the Shanghai Science and Technology Committee (Grant No. 22JC1410300) and Shanghai Key Laboratory of Material Frontiers Research in Extreme Environments (Grant No.22dz2260800). This work at High Magnetic Field Laboratory was supported by the National Key R&D Program of the MOST of China (Grant No. 2022YFA1602602), the National Natural Science Foundation of China (Grant No. 12474053), the Basic Research Program of the Chinese Academy of Sciences Based on Major Scientific Infrastructures (Grants No. JZHKYPT-2021-08). Part of the transport measurements were performed at the High Magnetic Field Laboratory, CAS. The work at Fudan University was supported by the Key Program of the National Natural Science Foundation of China (Grant No. 12234006), the National Key R&D Program of China (Grant No. 2022YFA1403202), the Innovation Program for Quantum Science and Technology (Grant no. 2024ZD0300103). A portion of this research used resources at the beamline 17UM of the Shanghai synchrotron radiation facility, China.

**Author contributions**

H.M. and Q.Z. supervised the project; J.L.Z., Q.Z., and D.P. conceived the ideas and coordinated all the research efforts; L.C., Y.Z, E.Z. and J.Z. synthesized the single-crystal samples; D.P. performed the resistance measurements under pressure with the support of Q.Z.; D.P. and Z.X. conducted the susceptibility measurements under pressure with the support of Q.Z.; D.P., Z.X., F.L., Y.L., Y.W., C.W., Y.Y. H.D. and H.L. performed the synchrotron XRD measurements and analysis with the support of Z.Z. and Q.Z.; D.P. Y.B. Z.X. J.C. T.L. Z.W. carried out the transport measurements assisted by Y.P. and M.T. D.P. Y.B. Z.Q., and J.L.Z. analyzed the data; J.L.Z., D.P., and Q.Z. wrote the paper with input from all other authors. All authors provided comments on the paper.

**Competing interests**

The authors declare no competing interests.

**Data availability**

The data supporting the findings of this research are available within the article and its Supplementary Materials.

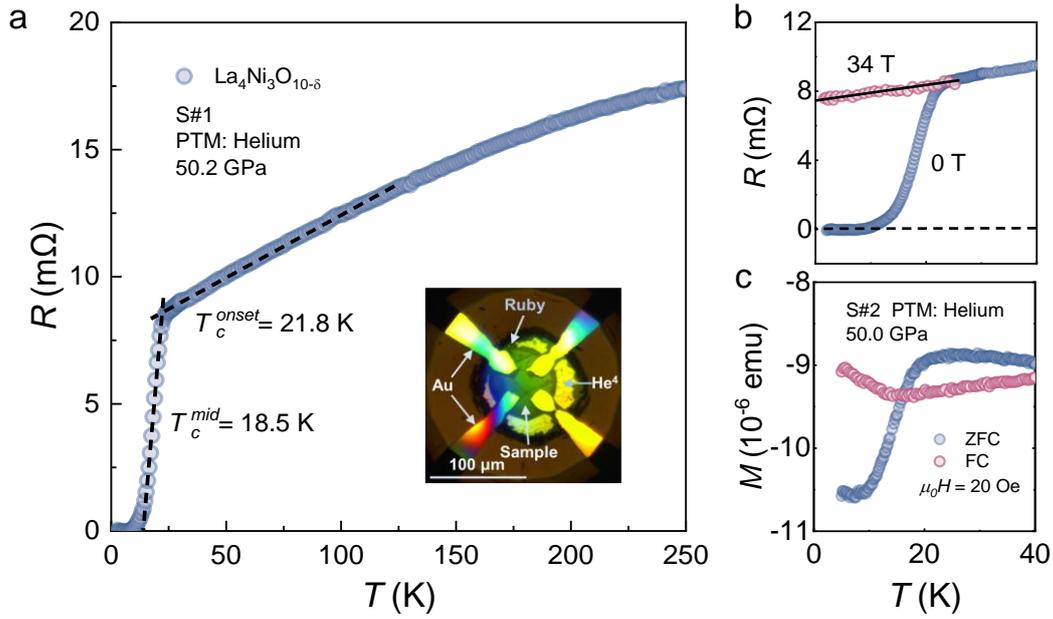

**Fig.1| Superconductivity of $La_4Ni_3O_{10-\delta}$ single crystal at ~50 GPa. a,** Temperature dependence of resistance of $La_4Ni_3O_{10-\delta}$(S#1) at 50.2 GPa. The superconducting transition temperature is determined from the resistance with onset at 21.8 K and midpoint at 18.5 K, respectively. The black dashed line demonstrates the linear-$T$ fit of the normal state resistance. Inset, a photograph of the electrodes used for high-pressure resistance measurements. The lamella was carved from a single crystal by using focused-ion-beam (FIB) milling and mounted with $ab$-plane parallel to the culet of DAC. The sample was loaded under helium as the pressure-transmitting medium. **b,** Temperature dependence of the resistance at zero field (blue dots) and high field (34 T, red dots). The black dashed line is a guide for T-linear resistance to the lowest temperature. **c,** The raw data of the DC magnetization for $La_4Ni_3O_{10-\delta}$(S#2) under a pressure of 50.0 GPa[6]. Magnetization measurements were performed in the nitrogen DAC, with a magnetic field of 20 Oe applied perpendicular to the $ab$-plane. A clear superconducting diamagnetic response is clearly observed in the ZFC curve.

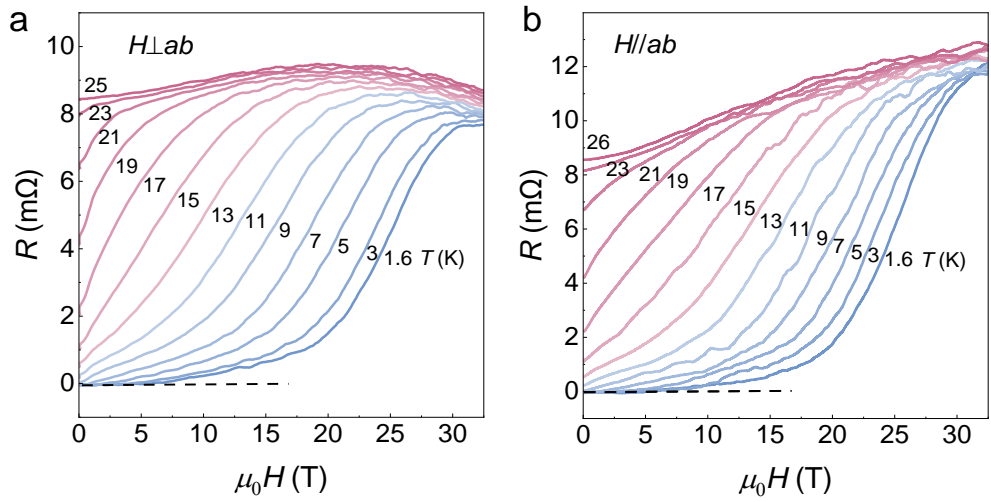

**Fig.2| Magnetic field dependence of the resistance of $La_4Ni_3O_{10-\delta}$ single crystal at 50.2 GPa.**
**a,b,** Magnetic field dependence of resistance at different temperatures with $H\perp ab$ (**a**) and $H//ab$ (**b**). The superconducting transition shifts to a lower magnetic field as temperature increases. The magnetoresistance effect is weak for $H\perp ab$, while it is pronounced in the case of $H//ab$. As a consequence, in the presence of an in-plane magnetic field, the SC transition becomes significantly broadened at higher temperatures.

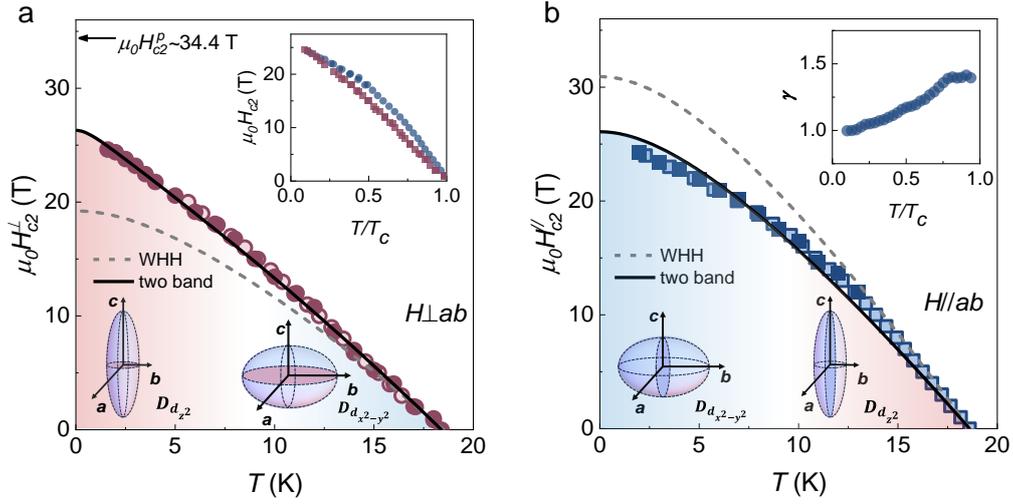

**Fig.3| The upper critical field $\mu_0 H_{c2}$ of La$_4$Ni$_3$O$_{10-\delta}$ single crystal at 50.2 GPa. a,b,** $\mu_0 H_{c2} - T$ phase diagrams for magnetic fields along the *c* axis (**a**) and in the *ab* plane (**b**). Solid circles (●) and squares (■) are obtained from the *R-H* curves (Figs.2**a,b**). Open circles (○) and squares (□) are taken from *R-T* curves (Extended Data Fig. 2). To minimize the effects of magnetoresistance and superconducting fluctuations, the values of $\mu_0 H_{c2}$ and $T_c$ are determined from the mid-point of the superconducting transitions. The black solid line represents the best fit of the two-band model in the case of strong interband coupling with the $\lambda_{11}=\lambda_{22}=0.2$ and $\lambda_{12}=\lambda_{12}=0.7$. The schematic diagrams illustrate the diffusivity of the bands primarily originating from $d_{z^2}$ and $d_{x^2-y^2}$ orbitals. The red and blue regions represent the behavior of $\mu_0 H_{c2}$ is determined by $D_{d_{x^2-y^2}}$ and $D_{d_{z^2}}$, respectively. The grey dashed line is the WHH fit without considering the spin paramagnetic effect. The inset of **a** overlays the $\mu_0 H_{c2}$ data for both field orientations. The inset of **b** shows the corresponding temperature dependence of the anisotropic parameter $\gamma = \mu_0 H_{c2}^{\parallel}/\mu_0 H_{c2}^{\perp}$.

**Synchrotron X-ray single crystal diffraction measurements at 50.2 GPa**

We performed synchrotron X-ray single-crystal diffraction measurements on sample S#3 at 50.2 GPa. As shown in Extended Data Fig. 1, the diffraction spots indicate that the sample maintains its single-crystal structure up to 50.2 GPa.

**Temperature dependence of resistance for La$_4$Ni$_3$O$_{10-\delta}$ (S#1) at specific magnetic fields.**

Extended Data Fig. 2 illustrates the temperature dependence of resistance for La$_4$Ni$_3$O$_{10-\delta}$ (S#1) under various magnetic fields. The symbols of the same color are extracted from the *R-H* curves at specific temperatures (Fig. 2a, b). The critical temperature is defined as the intersection point between the extrapolated resistance in the normal state and a line with the slope corresponding to the resistive transition at its midpoint. To accurately construct the *B-T* phase diagram for La$_4$Ni$_3$O$_{10-\delta}$ (S#1), the solid symbols in Fig. 3 are obtained from the *R-H* curves, while the open circles in Fig. 3 represent the critical temperatures derived from the *R-T* curves.

**Reproducibility**

To confirm the measurement results presented in the main text, we conducted additional magneto-transport measurements on a different sample (S#4) from the same batch. For S#4, the measurements were performed under a pressure of 48.6 GPa. As shown in Extended Data Fig. 3, a clear superconducting transition is observed for S#4, with an onset at 24.6 K and a midpoint at 21.7 K.

The magnetic-field dependence of the electrical resistance for S#4 is presented in Extended Data Fig. 4. The magnetoresistance effect is more pronounced in S#4; however, the shift in the resistive transition is qualitatively similar to that observed in S#1. In Extended Data Fig. 5, we show the temperature dependence of $\mu_0 H_{c2}(T)$ for both field directions, where the critical fields are determined from the midpoint of the superconducting transitions. The $\mu_0 H_{c2}(T)$ of S#4 also exhibits nearly isotropic behavior across the entire temperature range, with the anisotropy parameter $\gamma$ decreasing from 1.4 to 1.03 as the temperature decreases from $T_c$ to 1.8 K. Both $\mu_0 H_{c2}^{\perp}(T)$ and $\mu_0 H_{c2}^{\parallel}(T)$ can be well fitted using a two-band model. The fitting parameters are summarized in Table S1.

**Two-band model**

In the two-band model, the upper critical field can be described by a parametric equation:

$$a_0[ln(t) + U(h)][ln(t) + U(\eta h)] + a_1[ln(t) + U(h)] + a_2[ln(t) + U(\eta h)] = 0,$$

where $a_0$, $a_1$ and $a_2$ are determined by the intraband coupling constant ($\lambda_{11}$ and $\lambda_{22}$) and interband coupling constant ($\lambda_{12}$ and $\lambda_{21}$) where, $a_0 = 2(\lambda_{11}\lambda_{22} - \lambda_{12}\lambda_{21})/\lambda_0$, $a_1 = 1 + (\lambda_{11} - \lambda_{22})/\lambda_0$, $a_2 = 1 - (\lambda_{11} - \lambda_{12})/\lambda_0$ and $\lambda_0 = \sqrt{(\lambda_{11} - \lambda_{22})^2 + 4\lambda_{12}\lambda_{21}}$. $U(x) = \psi\left(x + \frac{1}{2}\right) - \psi\left(\frac{1}{2}\right)$, with $\psi(x)$ being the digamma function, where $T = T/T_c$, $h = \frac{\hbar H_{c2} D_1}{2\phi_0 T k_B}$, $\eta = D_2/D_1$, $D_1$ and $D_2$ are the diffusion coefficients of each band.

**Calculated $\mu_0 H_{c2}$ based on strong intraband coupling and strong interband coupling**

In Extended Data Fig. 6, we compare the calculated upper critical field in the case of strong intraband coupling $(\lambda_{12}\lambda_{21} \ll \lambda_{11}\lambda_{22})$ and strong interband coupling $(\lambda_{12}\lambda_{21} \gg \lambda_{11}\lambda_{22})$. For $\mu_0 H_{c2}^{\parallel}(T)$, due to the relatively large $\eta$, the calculated results for both cases show no significant difference. Our discussion primarily focuses on $\mu_0 H_{c2}^{\perp}(T)$. The $D_1$ can be estimated from $D_1 \approx \frac{8\phi_0 k_B}{\pi^2} / \frac{dH_{c2}}{dT}\big|_{T=T_c}$, when $D_1 \gg D_2$. The black solid line represents the calculated $\mu_0 H_{c2}$ for the case of strong interband coupling, with $\lambda_{11} = \lambda_{22} = 0.2$ and $\lambda_{12} = \lambda_{21} = 0.7$. The red solid line represents the scenario of strong intraband coupling, where $\lambda_{11} = 0.7$, $\lambda_{22} = 0.3$ and $\lambda_{12} = \lambda_{21} = 0.2$. Here, we took $D_1 = 1.2\ cm^2/s$. As shown in Extended Data Fig. 6, for the strong intraband coupling case, the calculated $\mu_0 H_{c2}$ exhibits a low-temperature upturn that deviates from the experimental data. This upturn feature becomes weaker with a further reduction of $\eta$, but the magnitude of the theoretical $\mu_0 H_{c2}$ would be much lower than the experimental data. **In the case of strong intraband coupling, even if we set $\eta$ as the free parameters, the theoretical $\mu_0 H_{c2}$ fails to describe the experimental data.**

**Diffusion coefficient of each band**

Generally, the upper critical field near $T_c$ is determined by the band with the higher diffusivity $D_1$. The band with lower diffusivity $D_2$ dominates the behavior of $\mu_0 H_{c2}(T)$ at low temperatures. The discussion of the diffusion coefficient for each band is based on the scenario of strong interband coupling. By fitting $\mu_0 H_{c2}^{\perp}(T)$, we can get an in-plane diffusion coefficient $D_1^{\parallel} = 1.2\ cm^2/sec$ and $D_2^{\parallel} = 0.09\ cm^2/sec$. The out-of-plane diffusion coefficient $D_1^{\perp} = 0.71\ cm^2/sec$ and $D_2^{\perp} = 0.39\ cm^2/sec$ can be determined from $\mu_0 H_{c2}^{\parallel}(T)$. The fitting parameters for S#1 and S#4 are summarized in Table S1.

Extended Data for Isotropic superconductivity in trilayer Nickelate $La_4Ni_3O_{10-\delta}$

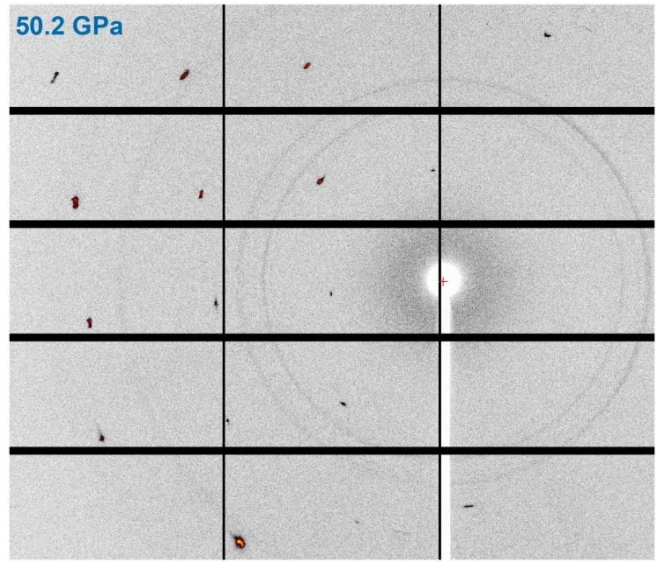

**Extended Data Fig. 1|: The single-crystal synchrotron X-ray diffraction pattern of the sample in a diamond anvil cell at 50.2 GPa.** The single-crystal featured diffraction pattern indicates the sample retains its single-crystal structure up to 50.2 GPa. The extra power rings are from the rhenium gasket due to the relatively large tail of the x-ray beam.

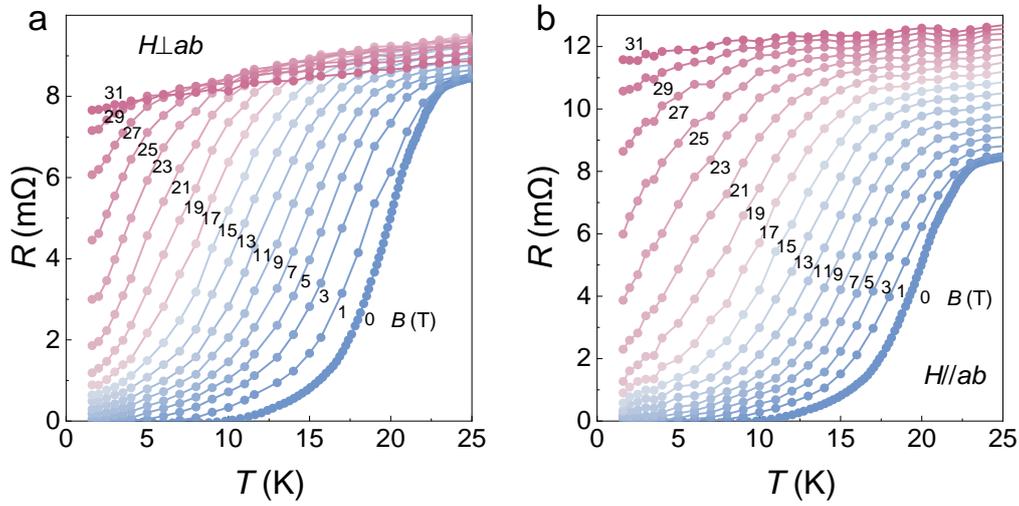

**Extended Data Fig. 2|: Temperature dependence of resistance for $La_4Ni_3O_{10-\delta}$ (S#1) at specific magnetic fields.** Resistance versus temperature at selected magnetic fields, $H\perp ab$ (**a**) and $H//ab$ (**b**). Symbols with the same color are determined from the *R-H* curve at a specific magnetic field (**Fig.2 a,b**). The solid lines are guides to the eye.

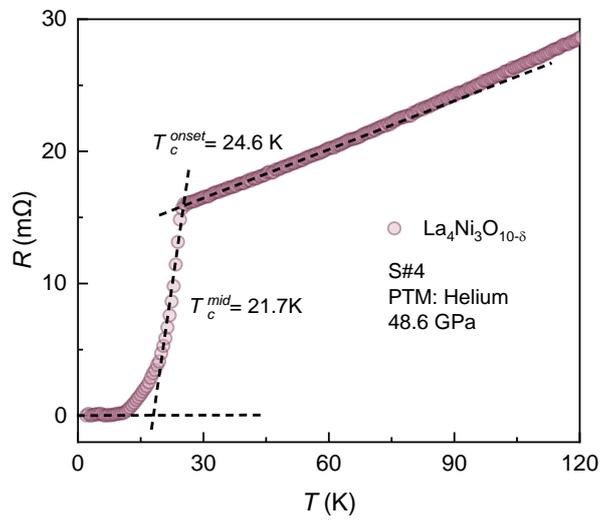

**Extended Data Fig. 3|: Superconductivity in $La_4Ni_3O_{10-\delta}$ (S#4) at 48.6 GPa.** Temperature dependence of the electrical resistance for S#4.

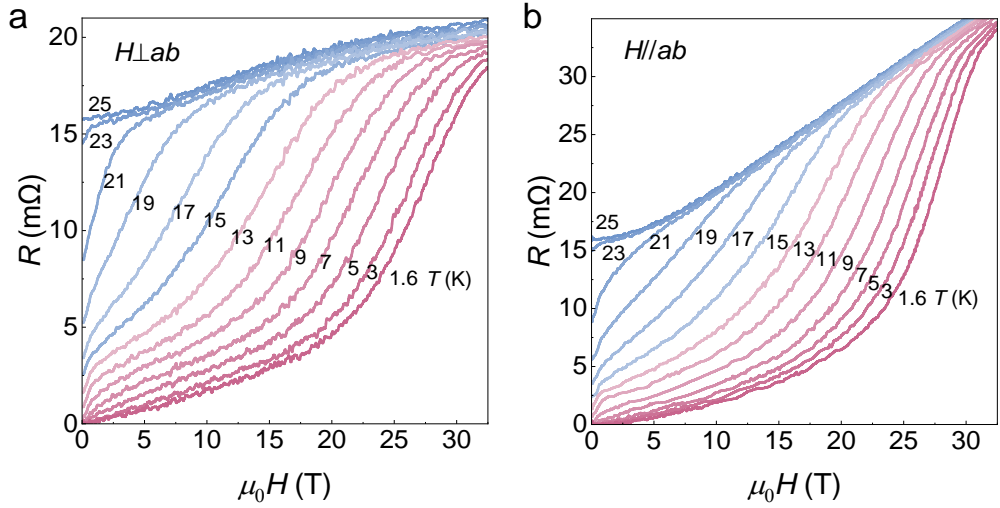

**Extended Data Fig. 4|: Magnetic field dependence of the resistance of La$_4$Ni$_3$O$_{10-\delta}$ (S#4) under a pressure of 48.6 GPa.** Magnetic field dependence of the electrical resistivity at variant temperatures for S#4: (a) $H\perp ab$ ; (b) $H//ab$.

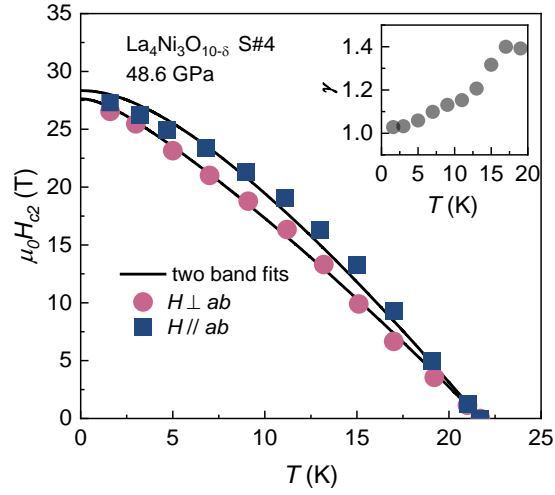

**Extended Data Fig. 5|: The upper critical field $\mu_0 H_{c2}$ of La$_4$Ni$_3$O$_{10-\delta}$ (S#4) under a pressure of 48.6 GPa.** Upper critical field versus temperature for (a) $H\perp ab$ ; (b) $H//ab$ as determined by a criterion of 50% of the normal-state resistivity. The solid lines are the best fits to the experimental data with two-band model. The inset shows the temperature dependence of the anisotropic parameter $\gamma = \mu_0 H_{c2}^{\parallel}/\mu_0 H_{c2}^{\perp}$.

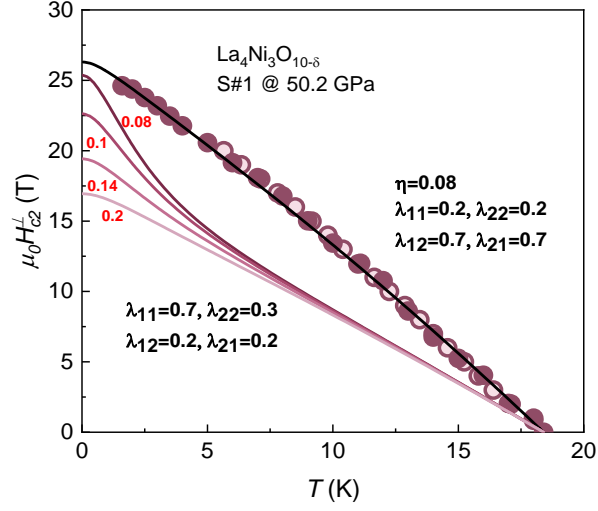

**Extended Data Fig. 6|: Analysis of the upper critical field based on two-band model**: Two-band fits to the out-of-plane upper critical field ($\mu_0 H_{c2}^{\perp}(T)$) for different scenarios. The black solid line represents the fitting with strong interband coupling, where $\lambda_{11}=\lambda_{22}=$**0.2**, $\lambda_{12}=\lambda_{12}=$0.7 and $\eta$=0.08. The red solid line represents the fitting with strong intraband coupling, where $\lambda_{11}$=0.7, $\lambda_{22}$=**0.3,** $\lambda_{12}=\lambda_{12}$=0.2, and we take $\eta$ from 0.08 to 0.2.

| Sample | pressure | $T_c$ | $\mu_0 H_{c2}^P(0)$ | Field direction | $\mu_0 H_{c2}$ (1.8 K) | $\frac{d\mu_0 H_{c2}}{dT}\big|_{T=T_c}$ | Two band fits | $\mu_0 H_{c2}^{orb}$ |
|---|---|---|---|---|---|---|---|---|
| **S#1** | 50.2 GPa | 18.5 K | 34.4 T | H⊥ab | 24.4 T | 1.5 T/K | $D_1^{\parallel}$=1.2 $cm^2/s$<br>$D_2^{\parallel}$=0.09 $cm^2/s$ | 19.2 T |
|  |  |  |  | H//ab | 24.4 T | 2.5 T/K | $D_1^{\perp}$=0.71 $cm^2/s$<br>$D_2^{\perp}$=0.39 $cm^2/s$ | 30.9 T |
| **S#4** | 48.6 GPa | 21.7 K | 40.3 T | H⊥ab | 26.6 T | 1.4 T/K | $D_1^{\parallel}$=1.18 $cm^2/s$<br>$D_2^{\parallel}$=0.18 $cm^2/s$ | 21.0 T |
|  |  |  |  | H//ab | 27.3 T | 2.2 T/K | $D_1^{\perp}$=0.75 $cm^2/s$<br>$D_2^{\perp}$=0.43 $cm^2/s$ | 33.1 T |

**Extended Data Table 1|:** Sample properties and fitted results.